\def\year{2020}\relax
\documentclass[letterpaper]{article} 
\usepackage{aaai20}  
\usepackage{times}  
\usepackage{helvet} 
\usepackage{courier}  
\usepackage[hyphens]{url}  
\usepackage{graphicx} 
\usepackage{graphicx}  
\nocopyright

\usepackage{times}
\usepackage{subfigure}
\usepackage{soul}
\usepackage{url}
\usepackage[utf8]{inputenc}
\usepackage[small]{caption}
\usepackage{graphicx}
\usepackage{amsmath}
\usepackage{amsfonts}
\usepackage{booktabs}
\usepackage{algorithm}
\usepackage{algorithmic}
\usepackage{comment}
\usepackage{url}
\usepackage{scalefnt}
\usepackage{bm}
\usepackage{multirow}
\usepackage{hyperref}
\hypersetup{colorlinks,allcolors=black}

\title{Fake News Detection using Temporal Features Extracted via Point Process}
\author{Taichi Murayama, Shoko Wakamiya, Eiji Aramaki\\
Nara Institute of Science and Technology\\
\{murayama, wakamiya, aramaki\}@is.naist.jp}

\begin{document}
\maketitle

\begin{abstract}
Many people use social networking services (SNSs) to easily access various news. 
There are numerous ways to obtain and share ``fake news,'' which are news carrying false information.
To address fake news, several studies have been conducted for detecting fake news by using SNS-extracted features.
In this study, we attempt to use temporal features generated from SNS posts by using a point process algorithm to identify fake news from real news.
Temporal features in fake news detection have the advantage of robustness over existing features because it has minimal dependence on fake news propagators.
Further, we propose a novel multi-modal attention-based method, which includes linguistic and user features alongside temporal features, for detecting fake news from SNS posts.
Results obtained from three public datasets indicate that the proposed model achieves better performance compared to existing methods and demonstrate the effectiveness of temporal features for fake news detection.
\end{abstract}

\section{Introduction}
Social networking services (SNSs), such as Facebook and Twitter, provide many people with instant and convenient access to news.
However, SNSs constitute an effective platform for obtaining and sharing news that are not carefully fact-checked and may include false or uncertain information, called ``fake news.''
~\cite{fake_survey2} define ``fake news'' as ``a news article or message published and propagated through media, carrying false information regardless the means and motives behind it.''
In our paper, the same definition is used.

The wide spread of fake news cannot only harm social media platforms but society in general.
For example, during the US 2016 presidential election, fake news favoring different candidate were shared more than 37 million times on SNSs and strongly affected the election results ~\cite{2016election1,2016election2}.
Consequently, the unprecedented growth of fake news reflects a strong need for detecting and mitigating fake news circulation ~\cite{2018_sciencefake}.
To confront these societal challenges, websites such as Snopes.com\footnote{https://www.snopes.com/} and PolitiFact.com\footnote{https://www.politifact.com/} track and debunk rumors and manually assess rumor credibility based on evidence.
These fact-checking sites are expensive to operate legitimately and require a considerable amount of time to validate and publish the credibility of a rumor.
Contrary to fact-checking websites, existing work on fake news detection mainly applies machine-learning methods based on various characteristics of SNSs, e.g., text content~\cite{rnn_fake}, user characteristics~\cite{user_fake} and propagation paths/trees~\cite{prop2}.

In addition to existing features, it is assumed that the temporal movements of SNS posts are also useful for detecting fake news~\cite{icwsm2020}.
Recent research~\cite{bots} showed that social bots influence the initial spread of fake news. 
Time series of posts referring to fake news exhibit different movement from those of real news.
Nevertheless, few studies have considered the amount of attention fake news attract over time.

This study proposes a fake news detection model that takes advantage of the attention to news changing over time, i.e., the temporal features.
The attention is calculated using a self-exciting point process from the post publication time and the likelihood of people reading the post (determined by the number followers).
In this study, we designate the attention to the news as an ``infectiousness value'' because it can be measured based on the probability of re-share of the information by each new user. 
The infectiousness value can be regarded as an index of the public interest in the news and, for real news, it normally decreases over time.
Conversely, our underlying assumption is that the infectiousness value of fake news upsurges twice: the first upsurge results from the original news (including the false information), and the second results from news items for which people doubt or correct the false information.

The infectiousness value of the information is more robust than that of existing features, which depend on fake news propagators.
For example, text features of early users can be easily manipulated by providing fake comments for diffusion. 
User features and user-article relationship are being transformed by the regulation of platforms and account suspension. 
Propagation paths/trees are difficult to manipulate but it is expensive to obtain them.
Infectiousness values are also difficult to manipulate because the values are calculated from a series of posts, not by early movement.
Furthermore, the number of followers and post publication time, which are used for calculating the infectiousness values, can be easily obtained.

The proposed fake news detection model leverages three features: combing existing features, texts, and users with an Attention-based mechanism and implementing the infectiousness value.
As preliminary research, we investigate whether temporal features can distinguish real news from fake news to validate their effectiveness.
Then, experiments are carried out to demonstrate that each module, such as the temporal features, is useful for detecting fake news.

The contributions of this study are as follows. (1) We elucidate the differences of infectiousness values associated with real and fake news and consider the differences for fake news detection using a point process. 
(2) We propose a new multi-modal method that combines text and user features with infectiousness values.
(3) We show the effectiveness of the proposed model for fake news detection on SNSs through experimental procedures.

\section{Related Work}
Early studies attempted to detect fake news based on linguistic features extracted from texts~\cite{early_lang1,early_lang2}.
Recent studies used deep learning models to capture temporal--linguistic features.
~\cite{rnn_fake} used recurrent neural networks (RNNs), which capture temporal--linguistic features from a bag-of-words of user posts.
~\cite{rvnn_lang} used recursive neural networks based on the texts of a reply tree.
Further examples include convolutional neural networks ~\cite{cnn_lang}, hierarchical attention networks~\cite{hierarchical_lang}, and neural-network models using discourse-level structures~\cite{discourse_lang}.

Moreover, several methods were examined for detecting fake news using the characteristics of users who post the information.
In fact, ~\cite{early_lang1,user1,user_fake} used various models based on user characteristics, such as the number of followers, number of friends, and registered age.
Recently, the relationship between news articles and users is used to determine news credibility assuming that if a strong relation exists between two articles as determined by the number of users who re-shared them, the two articles are likely to share the same label~\cite{markov}.

Other studies employ detection methods based on propagation paths/trees or networks of posts on SNS.
~\cite{prop3} proposed a graph-kernel-based support vector machine (SVM) classifier that calculates the similarity between propagation tree structures.

Multi-modal approaches combine features of different types to detect fake news.
For example, ~\cite{multi2} combined texts and user behavior, while ~\cite{multi1} combined texts and visual features extracted from SNS posts.
Our model effectively combines text and user features using contextual inter-modal attention~\cite{CIM} to catch the relationship between a user and a post content.

In a method of fake news detection using temporal features similar to the proposed method, ~\cite{similar_3} demonstrated the importance of using post temporal information for rumor stance classification.
~\cite{similar_1} used SpikeM~\cite{spikeM} to mathematically capture the time series behavior of information for long-term rumor detection, in addition to using other features (e.g., linguistic, user, network). 
In this study, we demonstrate that temporal features are also useful for short-term fake news detection.
The proposed multi-modal framework utilizes linguistic, user, and temporal features, which are easy to obtain, to capture the characteristics of fake news.

\section{Preliminary Research}
\begin{figure}[t]
    \includegraphics[width=8.5cm]{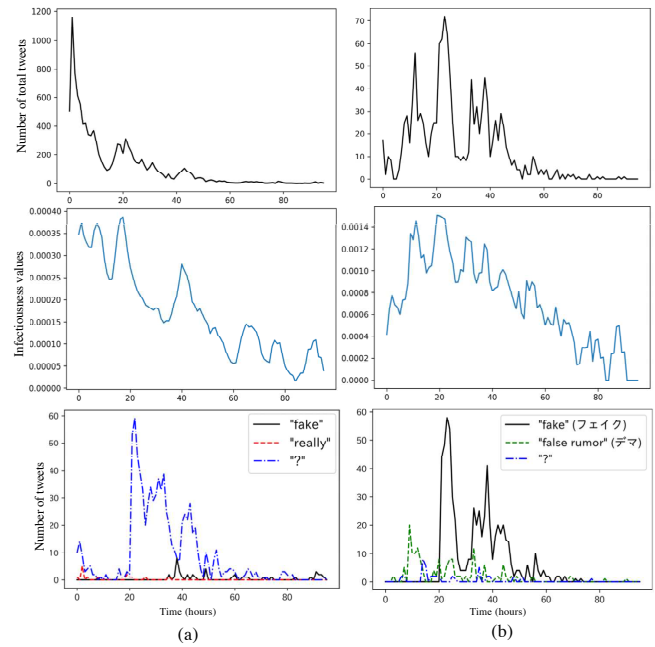}
    \caption{Time series of two fake news items regarding: (a) Islamic news in the U.S. and (b) Okinawa's news in Japan extracted from Twitter API for a 96-hour observation period. 
    The time series show the number of total tweets in the upper panels, the infectiousness values calculated using a self-exciting point process in the middle panels, and the number of tweets which include words used for doubts and denials (``fake,'' ``really'' and ``?'' as words for the U.S. and ``fake,'' ``false rumor'' and ``?'' as words of Japan), in the bottom panels.
    The result of (a) indicates few explicit words such as ``fake,'' but the question mark which represents doubt appeared many times in the same timing as the second upsurge around 20 hours. The result of (b) shows that explicit words indicating news as fake/false, appeared after the first upsurge at around 22 hours.}
    \label{case_word}
\end{figure}

We validated the contribution of temporal features in SNS posts to judge whether the news are fake or real (not fake).
Using Twitter API~\footnote{https://developer.twitter.com}, we obtained real and fake news items published in 2019 in the U.S. and Japan.
Additionally, for the U.S. news, we collected posts about fake news from the URLs and keywords, as extracted from Snopes.com and PolitiFact.com articles. 
Because Japan has no major fast-checking websites, for the Japan news, we collected posts about fake news from major media, public organizations, and companies denied in Japan.
We also collected real news from the URLs of news articles by major media.

Figure~\ref{case_word} presents the time series of the two fake news examples in the U.S. (a) and Japan (b).
Each news item has three time series.
The upper one indicates the number of tweets on each hour and the middle one indicates the infectiousness values calculated using the self-exciting point process described in the ``Fake News Detection Model'' section, which represents the probability of re-share.
It is thought that the time series of the number of tweets about real news shows a large upsurge in a few hours but decays quickly over time~\cite{seismic}.
Contrarily, the time series of the number of tweets about fake news (see upper panels) shows a second upsurge after approximately a day following the unstable behavior in the infectiousness value of fake news.
These behaviors are observed in other fake news and other countries.
An earlier study~\cite{similar_1} indicated that the time series of rumors have multiple upsurges during long-term observation periods (56 days), unlike those of non-rumors. 
In contrast, our results demonstrate that time series of fake news have multiple upsurges in short-term observations (4 days), unlike those of real news.
The time series graphs and a description of the collected news are presented in the URL~\footnote{\url{ https://docs.google.com/document/d/193Xv0AqmHB1F-UuaRuXpZOeMtfjNMnNrmTBUTjkoFIw/edit?usp=sharing}}.

We assumed that the multiple upsurges in the time series of fake news are caused by the attention received by posts questioning or denying the news.
To test this hypothesis, we examined whether the second upsurge coincides with the increase of posts expressing doubt and denial.
As shown in the bottom panels of Figure~\ref{case_word}, posts expressing doubt and denial appeared multiple times after the first upsurge within 48 hours.
For example, the result of (a) indicates few explicit words such as ``fake,'' but the question mark which represents doubt appeared many times in the same timing as the second upsurge around 20 hours. The result of (b) shows that explicit words indicating news as fake/false, appeared around 22 hours.
These results support our assumption and are mostly in agreement with a previous study~\cite{hoaxy}, which indicated a characteristic time lag between fake news and fact-checking.
Additionally, we have inferred that the multiple upsurges related to fake news are caused by renewed public interest because the meaning of news changes after questioning or denial (Fig.~\ref{case_word}).
The differences between the time series of fake and real news suggest that temporal features, which are more difficult to manipulate than others, can be useful for detecting fake news.

\section{Fake News Detection Model}

\begin{figure*}[t]
    \includegraphics[width=17cm]{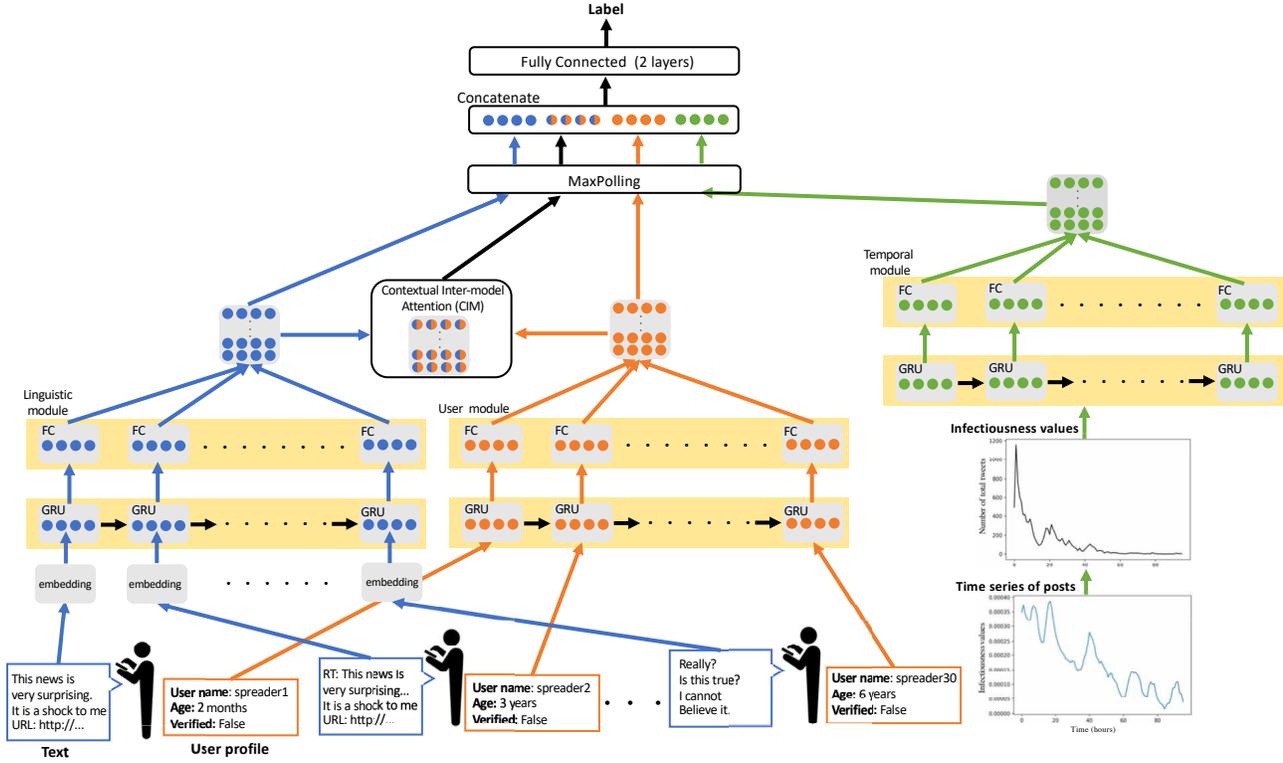}
    \caption{Architecture of the proposed fake news detection model. GRUs are used to learn the latent representations of linguistic, user, and temporal features. Then, CIM combines the linguistic and user features. Finally, the label of news is predicted by concatenating these features.}
    \label{model}
\end{figure*}

Although temporal features are useful, fake news detection using temporal features alone cannot achieve sufficient performance. 
Consequently, we propose a novel multi-modal method to detect fake news from many SNS posts.
The proposed model effectively combines linguistic and user features using an Attention module and then implements temporal features.
The overall model architecture is presented in Figure~\ref{model}.

\begin{table}[!t]
\center
\small
\caption{Major notations}
\begin{tabular}{c|l} \toprule
Notation & Definition \\ \bottomrule \toprule
$A_{i}$ & $i_{th}$ news story\\ 
$a_{t}$ & $t_{th}$ post of news story $A_{i}$\\ 
$\textbf{l}_{t}$ & linguistic feature of $t_{th}$ post\\
$\textbf{u}_{t}$ & user feature of $t_{th}$ post\\
$\textbf{s}_{i}$ & temporal features of $i_{th}$ news story\\
$\textbf{s}^{h}$ & infectiousness values at each point\\
$\textbf{l}^{e}_{t}$ & $I^{l}$-dimensional post embedding of $t_{th}$ post\\
\multirow{2}{*}{$\tilde{\textbf{h}}^{*}_{t}$, $\textbf{h}^{*}_{t}$} & hidden state of the $t_{th}$ post\\
& through GRU and FC in each module\\
$\textbf{h}^{max\_*}$ & each hidden states through MaxPooling\\
$\textbf{z}$ & the final output representing class probability\\
$H^{*}$ & each module output consisting of a sequence  [$\textbf{h}^{*}_{t}$]\\
$T^{*}$ & Number of sequence lengths of each module\\
\multirow{2}{*}{$E^{*}$} & Number of dimensions about hidden states $\textbf{h}^{*}_{t}$\\
& in each module\\
\bottomrule
\end{tabular}
\label{notation}
\end{table}

\subsection{Problem Statement}
The task of fake news detection is the prediction of the news label (real or fake), given the SNS posts related to the news.
Let $A_{i}$ be a news story consisting of $N_{i}$ posts; $A_{i} = \left\{a_{1}, a_{2},...,a_{N_{i}}\right\}$.
Each post $a_{t} = \left(\textbf{l}_{t}, \textbf{u}_{t}\right)$ consists of a linguistic feature $\textbf{l}_{t}$ and a user feature $\textbf{u}_{t}$.
The temporal features of a news story are represented as $\textbf{s}_{i}$.
Additionally, each news story $A_{i}$ is associated with a label $L\left(A_{i}\right)$, which has categorical variables $\left\{0, 1\right\}^{\tau}$.
We aim to learn a fake news detection function $f: f(A_{i}, \textbf{s}_{i}) \to L\left(A_{i}\right)$ that maximizes the prediction accuracy.

\subsection{Model Structure}
The model comprises various components.
The linguistic, user, and temporal modules convert inputs to latent features.
The contextual inter-modal attention module combines the latent features generated by the linguistic and user modules with attention.
Finally, the classification module outputs the prediction.
Table~\ref{notation} represents the major notations.

\subsubsection{Linguistic module}
We first converted the raw text of each post $a_{t}$ to the linguistic feature $\textbf{l}_{t}$ for model interpretation.
Then, we used the tf-idf values of the vocabulary terms of each post.
We used the top-$K$ vocabularies according to their tf-idf values.
Therefore, for each post, we extracted the linguistic feature $\textbf{l}_{t} \in \mathbb{R}^{K}$, which is a K-dimensional vector.
The linguistic feature $\textbf{l}_{t}$ created from the post corresponds to sparse high-dimensional data.
Therefore, we convert the vector $\textbf{l}_{t}$ into a low-dimensional representation.
Instead of using pre-trained vectors based on external collections, we learn the embedding matrix through our model;
$\textbf{l}^{e}_{t}=Embedding(\textbf{l}_{t})$, where $\textbf{l}^{e}_{t} \in \mathbb{R}^{I^{l}}$ denotes the $I^{l}$-dimensional post embedding vector of $\textbf{l}_{t}$.

From each post consisting of a sequence of embedded posts $L^{e}_{i} = \left[\textbf{l}^{e}_{1}, \textbf{l}^{e}_{2}, ..., \textbf{l}^{e}_{T^{l}}\right]$, we extract the latent linguistic features to use gated recurrent units~\cite{gru} (GRUs).
Actually, GRUs based on an RNN can capture long-term dependencies to learn the temporal--linguistic features of early posts on SNS.
A GRU takes $\textbf{l}^{e}_{t}$ and $\tilde{\textbf{h}}_{t-1}$ as input and produces $\tilde{\textbf{h}}_{t}$ as output. 
The respective formulas are described below:
\[
    \begin{aligned}
        \textbf{z}^{l}_{t} &= \sigma\left(U^{l}_{z}\textbf{l}^{e}_{t} + W^{l}_{z}\tilde{\textbf{h}}^{l}_{t-1}\right),\, \textbf{f}^{l}_{t} = tanh\left(U^{l}_{h}\textbf{l}^{e}_{t} + \tilde{\textbf{h}}^{l}_{t-1} \odot W^{l}_{h}\textbf{r}_{t}\right)\\
        \textbf{r}^{l}_{t} &= \sigma\left(U^{l}_{r}\textbf{l}^{e}_{t} + W^{l}_{r}\tilde{\textbf{h}}^{l}_{t-1}\right), \, \tilde{\textbf{h}}^{l}_{t} = \left(1 - \textbf{z}^{l}_{t}\right) \odot \tilde{\textbf{h}}^{l}_{t-1} + \textbf{z}^{l}_{t} \odot \textbf{f}^{l}_{t}\\
    \end{aligned}
\]
where $\textbf{z}^{l}_{t}$ and $\textbf{r}^{l}_{t}$ represent the reset and update gate at time $t$, respectively.
Furthermore, $U^{l}_{z}, U^{l}_{r}, U^{l}_{h} \in \mathbb{R}^{I^{l} \times E^{l}}$, $W^{l}_{z}, W^{l}_{r}, W^{l}_{h} \in \mathbb{R}^{E^{l} \times E^{l}}$ are parameters used for the respective gates.
$E^{l}$ denotes the output dimension of the GRU.
Then, the hidden state $\tilde{\bf{h}}^{l}_{t}$ of the GRU is applied by the fully connected (FC) layer, resulting in $\textbf{h}^{l}_{t} \in \mathbb{R}^{E^{l}}$, as shown below:
\begin{eqnarray}
    \tilde{\textbf{h}}^{l}_{t} = GRU\left(\textbf{l}^{e}_{t}\right), \quad \textbf{h}^{l}_{t} = FC(\tilde{\textbf{h}}^{l}_{t}), \quad t \in \left\{1, ..., T^{l}\right\}
\end{eqnarray}

\subsubsection{User module}
We used eight common characteristics extracted from SNS user profiles as the user features; the length of user description, length of user name, number of followers, number of follows, number of posts, registration age, and whether verified mark and geo information are attached to the account.
These are similar to ~\cite{user_fake}. 
The eight common features for a post $a_{t}$ are represented by $\textbf{u}_{t} \in \mathbb{R}^{I^{u}}$.
As with the linguistic features, we use GRUs to capture long-term dependencies and FC layers for the user features, as shown below:
\begin{eqnarray}
    \begin{aligned}
    \tilde{\textbf{h}}^{u}_{t} &= GRU(\textbf{u}_{t}), \quad \textbf{h}^{u}_{t} &= FC(\tilde{\textbf{h}}^{u}_{t}), \quad t \in \left\{1, ..., T^{u}\right\} \\
    \end{aligned}
\end{eqnarray}

\subsubsection{Temporal module}
In the previous section, we described the differences between the appearance time of posts about real and fake news.
To capture the potential components of this behavior, we convert the time series of posts to infectiousness values, which represent the re-share probability and drop as the news gets stale, via a self-exciting point process model (designated as SEISMIC)~\cite{seismic}.
SEISMIC, based on the Hawkes process~\cite{hawkes}, calculates the infectiousness value $s^{h}_{t}$ at time $t$ using the number of posts $R_{t}$ until time $t$ and the intensity $\lambda_{t}$.
$s^{h}_{t}$ is the input of the GRUs in the temporal module.
\begin{eqnarray}
    \lambda_{t} = s^{h}_{t}\sum_{t_{i}\leq t, i\geq 0}n_{i}\phi\left(t - t_{i}\right), \quad t\geq t_{0}. \\
    \phi\left(s\right) = 
    \begin{cases}
        c & if \quad 0 < s \leq s_{0}, \\
        c(s/s_{0})^{-\left(1+\theta\right)} & if \quad s > s_{0},
    \end{cases}
\end{eqnarray}
where $n_{i}$ represents the number of people accessing the news (number of followers).
Additionally, $\phi(\cdot)$ denotes the memory kernel, which quantifies the delay between the arrival and re-share of a post by a user. 
These parameters are estimated by ~\cite{seismic}: $s_{0}$ is 5 min, $\theta$ is $0.242$, and $c = 6.27 \times 10^{-4}$.
This process is designated as \textit{self-exciting} because each previous observation $i$ contributes to the intensity $\lambda_{t}$.

The estimation of the temporal variance of $s^{h}_{t}$ relies on a sequence of one-sided kernels $K_{t}\left(s\right)$, which up-weights the most recent posts and down-weights older posts.
These one-sided kernels keep the estimator $s^{h}_{t}$ close to the ever-changing real values.
\begin{eqnarray}
    s^{h}_{t} = \frac{\sum^{R_{t}}_{i=1}K_{t}\left(t - t_{i}\right)}{\sum^{R_{t}}_{i=0}n_{i}\int_{t_{i}}^t K_{t}\left(t - s\right)\phi\left(s - t_{i}\right) ds} \\
    K_{t}\left(s\right) = \max\left\{1 - \frac{2s}{t}, 0\right\}, \quad s > 0.
\end{eqnarray}
Eqs. (5) and (6) are used to calculate the infectiousness values $s^{h}_{t}$ from the publication time and number of followers of each post up to time $t$.
As described herein, $\textbf{s}_{i} = \left\{...,\left(time_{t}, follower_{t}\right),...\right\}, t \in \left\{1,...,N\right\}$ is the input of the function to convert to the infectiousness values, where $time_{t}$ represents the time elapsed from the first post.
Then, $\textbf{s}_{i}$ is converted to the infectiousness values $\textbf{s}^{h} = \left(s^{h}_{t_{1}}, s^{h}_{t_{2}},..., s^{h}_{t_{T^{s}}}\right)$ at each point, e.g., every hour.
As with the linguistic and user features, we utilize the GRUs and FC layers for the temporal features that are converted from every post information $\textbf{s}_{i}$, as explained below:
\begin{eqnarray}
\begin{aligned}
    \tilde{\textbf{h}}^{s}_{t} &= GRU(s^{h}_{t}), \quad \textbf{h}^{s}_{t} = FC(\tilde{\textbf{h}}^{s}_{t}) \quad t \in \left\{1, ..., T^{s}\right\} \\
    \end{aligned}
\end{eqnarray}

\subsubsection{Contextual Inter-model Attention}
Each post comprises linguistic and user features, which often have mutual interdependence.
However, GRUs are unable to capture characteristics of their interdependence.
Therefore, we used a pairwise contextual inter-modal attention mechanism (designated as CIM)~\cite{CIM}, using the latent representations generated by the GRUs.

We compute the attention between the output of the linguistic features $H^{l} = [\textbf{h}^{l}_{1},\textbf{h}^{l}_{2},...,\textbf{h}^{l}_{T^{l}}] \in \mathbb{R}^{T^{l} \times E^{l}}$ and that of user features $H^{u} = [\textbf{h}^{u}_{1},\textbf{h}^{u}_{2},...,\textbf{h}^{u}_{T^{u}}] \in \mathbb{R}^{T^{u} \times E^{u}}$ to leverage the contextual information related to each post to detect fake news, where $E^{l} \equiv E^{u}$ and $T^{l} \equiv T^{u}$.
First, a pair of matching matrices $M_{1}$, $M_{2} \in \mathbb{R}^{T^{l} \times T^{u}}$ are computed as $M_{1} = H^{l} \cdot {H^{u}}^\mathsf{T}, M_{2} = H^{u} \cdot {H^{l}}^\mathsf{T}$.

Furthermore, we obtained the probability distribution scores $N_{1}, N_{2} \in \mathbb{R}^{T^{l} \times T^{u}}$ over the respective matching matrices $M_{1}$ and $M_{2}$ to compute the attention weights on contextual posts using a softmax function.
Then, we computed the modality-wise attentive representations.
\begin{eqnarray}
    \begin{aligned}
    N_{1}\left(i,j\right) = \frac{e^{M_{1}\left(i,j\right)}}{\sum^{T^{l}}_{k=1}e^{M_{1}\left(i,k\right)}}, \quad & for \: i,j = 1,...,T^{l} \\
    N_{2}\left(i,j\right) = \frac{e^{M_{2}\left(i,j\right)}}{\sum^{T^{l}}_{k=1}e^{M_{2}\left(i,k\right)}}, \quad & for \: i,j = 1,...,T^{l} \\
    O_{1} = N_{1} \cdot H^{u},\quad O_{2} = N_{2} \cdot H^{l}
    \end{aligned}
\end{eqnarray}
Finally, we computed the element-wise matrix multiplication for the attention to the important components. 
Then, we concatenated the calculation values $A_{1}$ and $A_{2}$ to obtain the attention representations between $H^{l}$ and $H^{u}$.
\begin{eqnarray}
    \begin{aligned}
        A_{1} = O_{1} \odot H^{l}, \quad A_{2} = O_{2} \odot H^{u}\\
        H^{ul} = concat\left[A_{1}, A_{2}\right] \in \mathbb{R}^{T^{l} \times 2E^{l}}
    \end{aligned}
\end{eqnarray}

\subsubsection{Classification module}
After obtaining the features through the modules, we applied them to MaxPooling and concatenated each feature into a single vector $\textbf{f}^{1} \in \mathbb{R}^{E^{l} + E^{u} + 2E^{l} + E^{s}}$,
\begin{eqnarray}
    \textbf{f}^{1} = concat[\textbf{h}^{max\_l}, \textbf{h}^{max\_u}, \textbf{h}^{max\_ul}, \textbf{h}^{max\_s}]
\end{eqnarray}
where $\textbf{h}^{max\_*}$ indicates hidden states $H^{*}$ through MaxPooling, i.e., $ \textbf{h}^{max\_l} = MaxPooling\left(H^{l}\right)$.

For predicting the class label for each news item, we used FC layers with an activation function, such as $ReLU$ that consists of two layers, to identify the complex relations between the respective features.
The final output $\textbf{z} \in \mathbb{R}^{\tau}$ represents the probability distribution over the set of $\tau$ classes through the softmax function.
\begin{eqnarray}
    \begin{aligned}
        \textbf{f}^{2} = ReLU\left(FC\left(\textbf{f}^{1}\right)\right) , \quad \textbf{z} = Softmax\left(FC\left(\textbf{f}^{2}\right)\right) \\
    \end{aligned}
\end{eqnarray}

\section{Experimental Procedure}
\subsection{Datasets}
To experimentally evaluate our model, we used three publicly available datasets: Weibo released by ~\cite{rnn_fake}, and Twitter15 and Twitter16 released by ~\cite{prop3}.
Each dataset of posts related to fake news was collected from the most popular social media platforms, i.e.,  Weibo~\footnote{\url{https://www.weibo.com}} in China and Twitter~\footnote{\url{https://twitter.com}} in the U.S.
The Weibo dataset is annotated with one of two class labels: ``true'' or ``fake.''
The Twitter datasets are annotated with one of four class labels: ``true,'' ``fake,'' ``unverified'' or ``debunking of fake.''
Table~\ref{dataset} presents a summary of the datasets.
It should be noted that the dataset size is smaller at the time of release because some SNS stories and posts cannot be acquired owing to changes in disclosure statements and post deletion.

For the experiments, we divided each dataset into training, validation, and test sets.
Each dataset was split following a ratio of 3:1 for acquiring the training and test sets, respectively.
A 15\% of the training set was held for the validation set.

\begin{table}[!tb]
    \centering
    \caption{Summary of datasets}
    \begin{tabular}{l|rrr} \toprule
    Dataset & Weibo & Twitter15 & Twitter16 \\\midrule
    No. of true news & 2351 & 371 & 204\\
    No. of fake news & 2313 & 363 & 205\\
    No. of unverified news & - & 373 & 205\\
    No. of debunking & - & 372 & 199\\ \midrule
    No. of training posts & 2973 & 942 & 517\\
    No. of validation posts & 525 & 167 & 97\\
    No. of test posts & 1166 & 370 & 204\\ \bottomrule
    \end{tabular}
    \label{dataset}
\end{table}

\subsection{Comparative Methods}
We made comparisons between the proposed model and the following existing baseline methods of fake news detection.
\begin{itemize}
    \item \textbf{SVM-TS}~\cite{comp1}: A linear SVM classifier that uses time-series to model the variation of social context features.
    This model also uses diffusion-based features, such as the average number of re-shares, in addition to linguistic and user features.
    \item \textbf{CSI}~\cite{multi2}: CSI is a hybrid deep-learning model that uses information from user texts, responses, and behaviors.
    This model calculates the source characteristic based on the user behavior, and classifies an article as fake or not.
    \item \textbf{GRU-2}~\cite{rnn_fake}: 
    GRU-2 is equipped with two GRU hidden layers and an embedding layer following the input layer for learning rumor representations by modeling the sequential structure of relevant posts over time.
    \item \textbf{PPC}~\cite{user_fake}: PPC is a time series classifier that incorporates both recurrent and convolutional networks, which respectively capture user characteristics along the propagation path.
    \item \textbf{Proposed (w/o CIM)}: This is the proposed model without the contextual inter-modal attention module used for validating the effectiveness of CIM.
    \item \textbf{Proposed (w/o time)}: This model comprises two features for learning; i.e., it uses linguistic and user features for validating the effectiveness of the temporal features.
    \item \textbf{Proposed (freq)}: This model replaces the infectiousness values with the number of posts during each period for validating the effectiveness of the infectiousness values.
\end{itemize}

\subsection{Experimental Settings}
Our model has been trained to minimize the binary/categorical loss function while predicting the class label of each news item in the training set.
During training, all model parameters were updated using gradient-based methods following the AdaDelta update rule.
Additionally, Dropout, for which the value was set to 0.5, was applied on hidden layers $\tilde{\textbf{h}}^{*}_{t}, \textbf{h}^{*}_{t}, \textbf{f}^{1}$, and $\textbf{f}^{2}$ to avoid overfitting.
The number of training epochs was set to 500. 
Early stopping was applied as the validation loss saturated for 10 epochs.

The network structure and hyper-parameters were set based on the validation set and on previous studies~\cite{rnn_fake,user_fake}.
We set 5,000 vocabularies as top-$K$ based on the tf-idf values as input to the linguistic module.
These tf-idf values were converted to embedding vectors with a dimension $I^{l}$ of 100.
$I^{u}$ was set to eight, as described in the user module in the ``Fake News Detection Model'' section.
The sequence lengths of the GRUs for the linguistic and user features, $T^{l}$ and $T^{u}$, were chosen as above 30 in the Weibo dataset and above 40 in the Twitter15 and Twitter16 datasets, based on the results of a previous study~\cite{user_fake}.
Namely, we used the first 30 or 40 posts in a story time sorted in ascending order as the input of $T^{l}$ and $T^{u}$.

In the case study, most time series of the number of fake news posts showed a second upsurge after approximately one day after post publication.
Therefore, we set the infectiousness values on the first two days with a length $T^{s}$ of 47 as the input of the GRUs for the temporal features $\textbf{s}^{h}$.
These 47 infectiousness values were calculated using all data from the point publication time up to at each hourly point; i.e., $s^{h}_{t_{3}}$ is calculated by all posts up to 3 hours elapsed from the post publication.

The output size of each GRU ($E^{l}$, $E^{u}$, and $E^{s}$) is selected from (16, 32, 64, and 128) and the hidden dimension of the output FC layer $\textbf{f}^{2}$ is selected from ($E^{con}, \frac{E^{con}}{2}, \frac{E^{con}}{4}$, and $\frac{E^{con}}{8}$) in the validation period, where $E^{con}$ is the size of $\textbf{f}^{1}$, equal to $(E^{l} + E^{u} + 2E^{l} + E^{s})$.

We used the accuracy and F1-measure as metrics to evaluate the model capabilities.
Classification tasks, such as fake news detection, are commonly evaluated by the accuracy while F1-measure works complementary to address class imbalance.
We used the accuracy over all categories and the F1-measure for each class to evaluate the model performance.

\section{Results and Discussion}

\begin{table*}[!tb]
  \centering
  \small
    \caption{Fake news detection results on each dataset}
    \begin{tabular}{l|ccc|ccccc|ccccc} \toprule
         Dataset & \multicolumn{3}{c|}{Weibo} & \multicolumn{5}{c|}{Twitter15} & \multicolumn{5}{c}{Twitter16} \\ \midrule
         \multirow{2}{*}{Method.} & \multirow{2}{*}{Acc.} & \multicolumn{2}{c|}{$F_{1}$} & \multirow{2}{*}{Acc.} & \multicolumn{4}{c|}{$F_{1}$} & \multirow{2}{*}{Acc.} & \multicolumn{4}{c}{$F_{1}$}\\ 
         & & T & F & & T & F & U & D & & T & F & U & D\\
         \midrule
        SVM-TS & 0.827 & 0.831 & 0.837 & 0.599 & 0.772 & 0.598 & 0.608 & 0.544 & 0.574 & 0.743 & 0.488 & 0.551 & 0.549\\
        CSI &  0.780 & 0.750 & 0.803 & 0.556 & 0.601 & 0.631 & 0.550 & 0.530 & 0.507 & 0.552 & 0.511 & 0.475 & 0.443\\
        GRU-2 & 0.876 & 0.872 & 0.879 & 0.794 & 0.822 & 0.815 & 0.849 & 0.697 & 0.750 & 0.761 & 0.750 & \textbf{0.771} & 0.723\\ 
        PPC & 0.914 & 0.912 & 0.917 & 0.806 & 0.748 & 0.840 & 0.807 & 0.730 & 0.778 & 0.803 & 0.760 & 0.711 & 0.767 \\\midrule
        Proposed (w/o CIM) & 0.920 & 0.922 & 0.917 & 0.814 & 0.807 & 0.813 & \textbf{0.870} & 0.745 & 0.791 & 0.850 & 0.782 & 0.747 & 0.791\\
        Proposed (w/o time) & 0.912 & 0.913 & 0.910 & 0.814 & 0.857 & 0.806 & 0.868 & 0.677 & 0.791 & 0.864 & 0.829 & 0.717 & 0.776\\
        Proposed (freq) & 0.921 & 0.931 & 0.908 & 0.807 & 0.872 & 0.815 & 0.828 & 0.660 & 0.805 & 0.864 & 0.801 & 0.740 & 0.699\\
        Proposed model & \textbf{0.937} & \textbf{0.937} & \textbf{0.936} & \textbf{0.831} & \textbf{0.880} & \textbf{0.850} & 0.833 & \textbf{0.758} & \textbf{0.819} & \textbf{0.870} & \textbf{0.831} & 0.739 & \textbf{0.841}\\ \bottomrule
  \end{tabular}
    \label{result}
\end{table*}

The experimental results are presented in Table~\ref{result} and indicate that the proposed model outperforms most baseline methods, confirming the benefits of the multi-modal method and temporal features.
The baseline \textbf{SVM-TS}, based on hand-crafted features, was a better model because it combined various features, including linguistic, user, and temporal features.
Contrarily, \textbf{CSI} achieved low accuracy. 
The model calculates the user relation score from the training data and then detects fake news from the test data by using the scores of users who appear in both training and test data.
Because few users appeared in both the training and test datasets in our experiments, CSI performed poorly.
Most deep learning-based models, such as the \textbf{Proposed model}, \textbf{GRU-2}, and \textbf{PPC}, outperformed feature engineering-based models, such as \textbf{SVM-TS}.
Deep neural networks helped to learn better hidden representations of people's responses to the news on SNS for fake news detection.
The results show that \textbf{GRU-2} and \textbf{PPC}, which used linguistic and user features, respectively, to capture complex hidden features indicative of the corresponding responses, achieved a high accuracy and high F1-measure.

To validate the effectiveness of each module, we also conducted experiments using models that excluded CIM and the temporal features of the proposed model.
Compared to \textbf{Proposed (w/o CIM)}, \textbf{Proposed model} achieved a higher accuracy and F1-measure on all datasets, except for the unverified label data.
This result demonstrates that it was insufficient to learn the hidden representations of the user and linguistic features differently.
Moreover, inter-dependencies between the linguistic and user features were useful to detect whether a news item was fake or not because posts consist of both features. 
Compared to \textbf{Proposed (w/o time)}, \textbf{Proposed model} achieved higher scores, except for the unverified label data in Twitter15.
In a previous study~\cite{similar_1}, the time series of rumors is useful to detect rumors in long-term observation periods (56 days).
However, these results support our claims that temporal features can be useful for short-term fake news detection (2 days).
\textbf{Proposed (freq)} replaced the infectiousness values with the number of posts in each period for validating the conversion to the infectiousness values.
Its accuracy was slightly higher than that of Proposed (w/o time) for the Weibo and Twitter16 datasets when the number of posts was added.
Simultaneously, the degree of increased accuracy was not significantly higher than that of the Proposed model.
This result shows that conversion to infectiousness values is useful to catch latent information from the temporal features for the fake news detection.

\textbf{Proposed model} overall performed the best for most measures and datasets, demonstrating the effectiveness of our model compared to baseline methods.
Specifically, our model achieved the highest accuracy for the Weibo test subset (0.937), the Twitter15 test subset (0.831), and the Twitter16 test subset (0.819).
Additionally, our model achieved the highest performance in terms of the F1 score on the True, Fake, and Debunking news data labels.
However, similarly to the compared methods, our model did not produce good results for classifying unverified labels.
Presumably, effectively classifying ambiguous labels, such as unverified, is challenging even when implementing the temporal features.

\begin{figure}[t]
    \includegraphics[width=8cm]{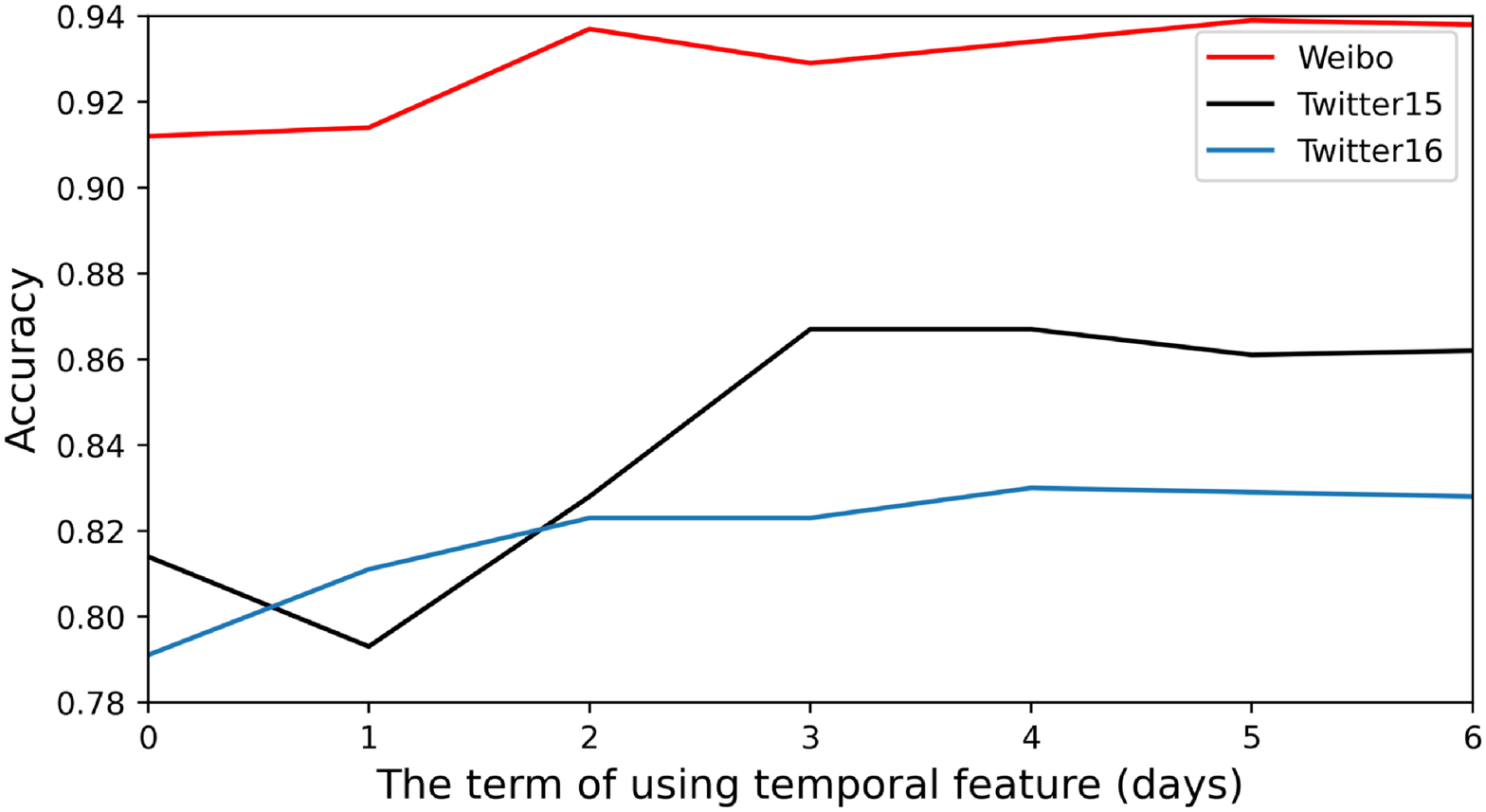}
    \caption{
    Accuracy of the proposed model with temporal features obtained for different time frames for each dataset: the horizontal axis represents the time frames from 0 (w/o time) to 6 days; the vertical axis represents the accuracy. 
    It is observed that better accuracy is achieved when a longer time frame is used.}
    \label{temporal}
\end{figure}

Finally, we evaluated the details of the contributions of the temporal features.
To examine the contributions, we compared the proposed models with varying time frames to obtain the temporal features from zero (w/o time) over six days (see Figure~\ref{temporal}).
The accuracy of the proposed model improves gradually as the time frame lengthens.
However, the proposed model performance remains more or less unchanged for a time frame over three days.
Specifically, the model accounting for five days of the Weibo dataset achieved an accuracy of 0.939.
When accounting for 4 days of the Twitter15 and Twitter16 datasets, our model achieved an accuracy of 0.867 and 0.830, respectively.
Although we set the time frame to the first 2 days in the experimental settings, the results show that time periods of approximately 4 or 5 days would be more appropriate for obtaining the temporal features for fake news detection.

Although our model demonstrates that incorporating temporal features, which are difficult to manipulate, in fake news detection models is useful, limitations also exist; it is difficult to detect ``early'' fake news.
The comparative method PPC claims to achieve fake news detection within an hour.
However, it is difficult to accurately estimate the infectiousness values $s^{h}_{t}$ of the information within an hour, so our model is not suitable for detecting early fake news.
Therefore, our results suggest the use of different models depending on the circumstances; models without temporal features are better for early detection, while the proposed model with temporal features are better for robust and high-precision detection.


\section{Conclusion}
We conclude this paper by highlighting the key points of our study:
(1) We ascertained the differences in time series behaviors between real and fake news from short-term observations.
(2) We proposed a novel multi-modal method for fake news detection, combining text and user features and infectiousness values.
(3) The experimental results empirically showed the effectiveness of the proposed model for the fake news detection problems.
However, it remains unclear whether the temporal features are useful in ambiguously labeled data (e.g., debunking label).
Future studies must examine how temporal features can be used flexibly effectively classifying ambiguous data labels.

\bibliographystyle{aaai}
\bibliography{ref}

\end{document}